\documentclass[rnote]{aa} 
\usepackage{graphicx}
\usepackage{float}
\usepackage{txfonts}
\usepackage{natbib}

\newcommand{\lsp}{LS~I~+61$^{\circ}$303}

\newcommand{\lsi}{LS~I~+61$^{\circ}$303~}

\newcommand{\beq}{\begin{equation}}
\newcommand{\eneq}{\end{equation}}

\begin{document}

\title{
Implications of the radio spectral index transition in \lsi for its INTEGRAL data analysis}

\author{L. Zimmermann
         \inst{1}
          \and
          M. Massi\inst{1}
          }

   \institute{Max Planck Institut f\"ur Radioastronomie, Auf dem H\"ugel
69, 53121
Bonn, Germany\\
              \email{lzimmerm@mpifr-bonn.mpg.de, mmassi@mpifr-bonn.mpg.de}
             }
   \date{Received 2011; }

\abstract
{The TeV emitting X-ray binary \lsi has two radio periodicities that correspond to a large periodic outburst with the same period as the orbit, 26.5 days (phase $\Phi$), and a second periodicity of 1667 days (phase $\Theta$), which modulates the orbital phase and amplitude of the large outburst. Analyses of the
radio spectral index revealed in \lsi the presence of the critical transition typical for microquasars from optically thick emission (related to a steady jet) to an optically thin outburst (related to a transient jet), and found that it occurs at $\Phi_{crit}$, which is modulated by $\Theta$: $\Phi_{crit}=f(\Theta)$.}
{We examine the possible implications of averaging high energy data over large $\Theta$ and $\Phi$ intervals in the light of puzzling published INTEGRAL results, which differ for different averaging of the data.}
{In microquasars, a simultaneous transition between two X-ray states occurs at the switch from optically thick radio emission to an optically thin radio outburst, from the low/hard to the steep power-law state. Assuming that the same transition occurs in \lsi at $\Phi_{crit}$, we can show qualitatively the effect of averaging high energy data on $\Theta$, by analysing the effects of averaging radio spectral index data across the same $\Theta$ interval. We then model the two X-ray states, low/hard and steep power-law state, and show quantitatively how their mixing can affect the results.}
{When folded over too large a $\Theta$ interval, spectral data from
INTEGRAL can yield a false picture of the emission behaviour of the source
along the orbit because it may be mixing two spectral states. Furthermore,
averaging the data along the orbit may result in a dominant low/hard
spectral state, which, for insufficiently extended sampling, might appear without a cut-off.}
{The INTEGRAL results can be interpreted as two X-ray states that alternate with each other along the orbit. The implications of this analysis for HE/VHE data from \lsi and a possible connection between HE/VHE emission and the steep power-law state are discussed.}


\titlerunning{Implications of the radio spectral index transition}

   \maketitle

\section{Introduction}
The clear periodicity of the X-ray binary  \lsp, formed by a compact object and a rapidly rotating B0 V star, is amongst its most peculiar characteristics. The orbital period of P = 26.496 days (phase $\Phi$ with $\Phi_{periastron}=$ 0.23-0.28, Aragona et al. 2009; Casares et al. 2005) modulates the radio flux resulting in a large outburst towards apastron, whereas a superorbital period of 1667 days (phase $\Theta$) modulates both the amplitude and the orbital phase of the large radio outburst \citep{Gregory99, Gregory02}. Two models exist for this system: 
the two-peak microquasar model \citep{MartiParedes, BoschRamon, Romero2007}
and the pulsar model \citep{Maraschi81,Dubus}. 
The name of the first model was coined because of the highly eccentric orbit in \lsp, for which two ejections are expected along the orbit: one around periastron and a second one shifted about $\Delta\Phi\sim 0.3$ towards apastron. In the radio regime, the first model interprets the often observed one-sided radio morphology and as well the observed switch to a double-sided morphology, as that of a precessing microquasar with a small and variable angle between the jet and the line of sight \citep{massi04,MassiZimmermann}. The second model interprets the one-sided structures as a cometary tail of the pulsar \citep{Dhawan06}.

\citet{MassiKaufman} analyzed the dependence of the radio spectral index $\alpha$ with flux density $S\propto\nu^{\alpha}$, on both periods $\Theta$ and $\Phi$. The analysis uses 6.7 years (1994-2000) of NASA/NRAO GBI data from \lsi at two frequencies, $\nu_1=$ 2.2 GHz and $\nu_2=$ 8.3 GHz. The analysis
had two main findings. The first was that in \lsi the optically thin outburst occurs after an interval of optically thick emission. In microquasars, the optically thin outburst is due to a transient jet consisting of shocks propagating in a pre-existing optically thick emitting steady jet  \citep{FenderBelloniGallo, Massi10a}. The second remarkable finding resulting from the radio spectral index analysis of \citet{MassiKaufman} was that in \lsi this switch from one kind of a jet (steady jet) to the other one (transient jet) 
occurs twice along the orbit at the orbital phases predicted by the two-peak microquasar model. One sees a spectral transition, optically thick/ optically thin emission, first around periastron at $\Phi_{crit,1}=0.33\pm 0.13$ and then again towards apastron at $\Phi_{crit,2}=0.7\pm 0.13$. Owing to the qualitative and quantitative agreement between radio data and  the two peak  microquasar model, the microquasar scenario is assumed here as a working hypothesis to explain the puzzling results obtained from INTEGRAL observations. We use the well-established relationship for microquasars between radio and X-ray states \citep{FenderBelloniGallo}.

The two radio states, the steady jet  ($\alpha \geq$0) and the transient jet ($\alpha < $0), should be directly linked to two spectrally distinct X-ray states in the unified model of X-ray states with radio jets: the low/hard X-ray state and the steep power-law state \citep{FenderBelloniGallo}. The X-ray states are clearly distinguishable by their energy spectra \citep{Grove,FenderBelloniGallo, McClintockRemillard06}. A steady jet ($\alpha\geq$ 0) occurs in the low/hard X-ray state and its X-ray spectrum is characterized by a power-law with photon index $\Gamma\approx$ 1.5-1.8 and a cut-off at high energies. The high energy cut-off is different for different sources, with some showing a folding energy $E_f\approx$115 keV \citep{Grove} and others, as GX 339-4, a folding energy $E_f$ of only 66 keV \citep{Caballero09}. In particular for \lsp, the very low X-ray luminosity, the observed optical modulation correlated with the radio outburst (see Massi \& Kaufman Bernad{\'o} 2009), and the softening associated to a decrease in X-ray flux \citep{Sidoli06}, which is consistent with accreting systems in quiescence \citep{Corbel06,Corbel08}, all point towards the source exhibiting a very low hard, nearly quiescent, X-ray state. Finally, the transient jet ($\alpha<$ 0) is associated with the steep power-law X-ray state with a black-body component (with kT$\sim$1 keV) and a power-law with a photon index of  $\Gamma\gtrsim$ 2.4 and no cut-off, thereby intrinsically allowing also for high energy emission.

Data from \lsi taken with the International Gamma-ray Astrophysics Laboratory (INTEGRAL/ISGRI ~20-500 keV, \lsi is detectable in the ~20-150 keV range), however, yielded surprising results, finding e.g. along most of the orbit a photon index only compatible with a low hard state, but without a cut-off. The photon index of the hard X-ray spectrum, furthermore, depended on how much data was averaged together and whether it was also averaged over the orbital period \citep{Chernyakova06,Zhang10}. Here, we aim to explain these results by using the relationship between radio and X-ray states that allows us to predict the kind of emission expected at soft gamma-rays based on the radio emission, or viceversa. In particular, in \lsp, the $\Theta$ and $\Phi$ periodicities allow us to test this relationship by using GBI data, even if no simultaneous observations are available (as is the case for INTEGRAL). From the radio spectral transitions seen in the GBI data one can infer which spectral transition is expected in X-rays and when. The combination of radio spectral index data from different $\Theta$ intervals can then help us to understand what happens when INTEGRAL data is averaged over large parts of $\Theta$.

\begin{figure}
   \centering
\includegraphics[width=.34\textheight]{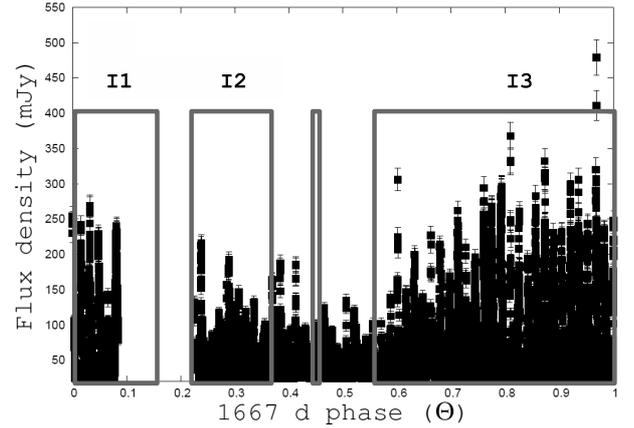}
      \caption{Published high energy observations of \lsi by INTEGRAL (I1,I2 and I3) shown in the context of the 1667 day radio period ($\Theta$). The underlying light curve gives 6.7 yr radio Green Bank Interferometer data at 8.3 GHz folded with  $\Theta$.}
         \label{Fig1}
   \end{figure}

\section{INTEGRAL observations versus $\Theta$ and $\Phi$}

In Fig. \ref{Fig1}, INTEGRAL observations published to date are displayed in terms of the $\Theta$ intervals in which they were carried out \citep{Zhang10} and denoted as I1, I2, and I3 (see also Fig. 1 in Zimmermann et al. 2011). Any spectral analysis should consider $\Theta$, because within the energy range covered by INTEGRAL data, two states, the low hard and the steep power-law state, can be detected, when they occur.

\begin{figure}
   \centering
\includegraphics[angle=-90, width=0.3\textwidth]{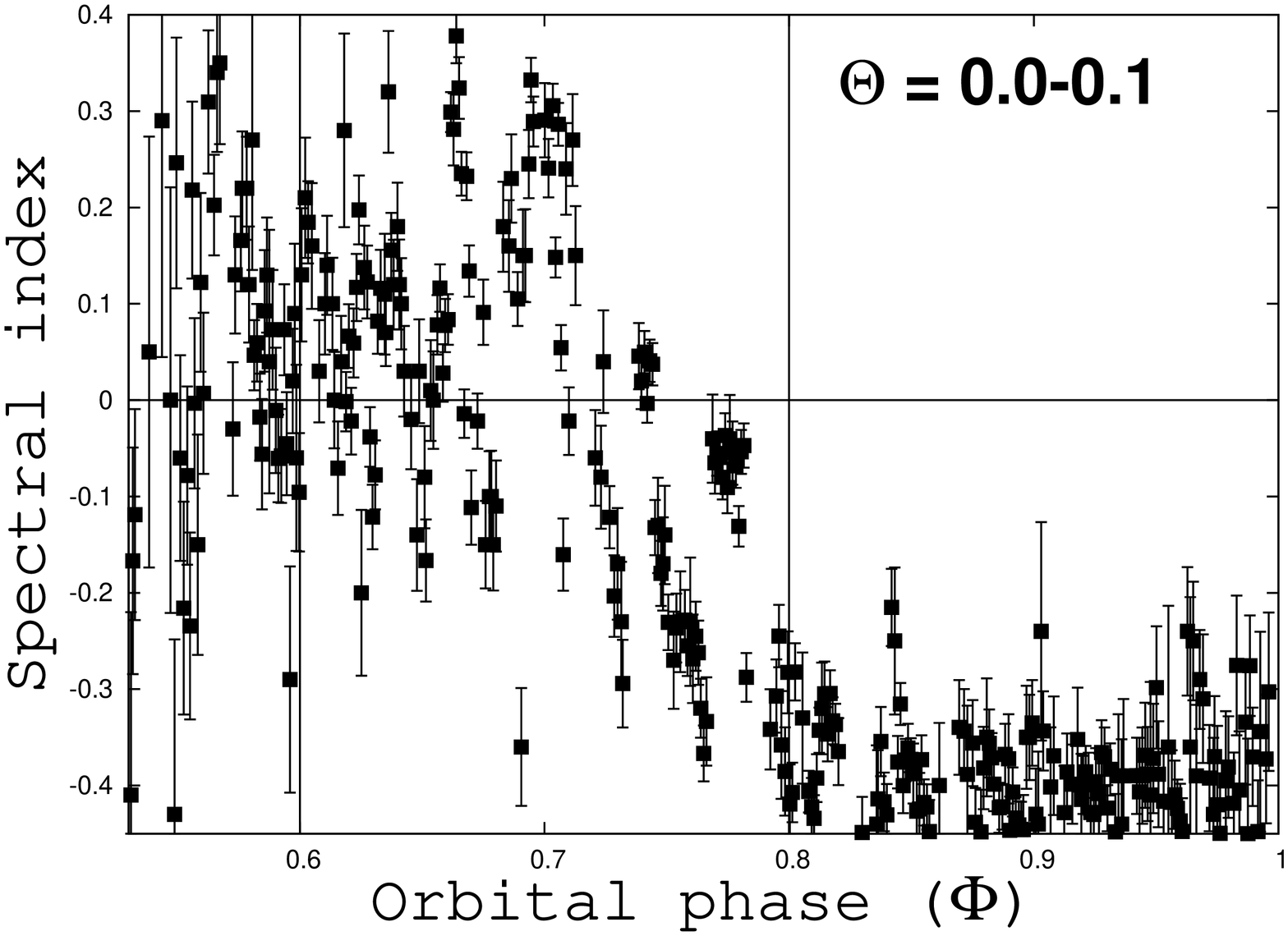}
\includegraphics[angle=-90, width=0.3\textwidth]{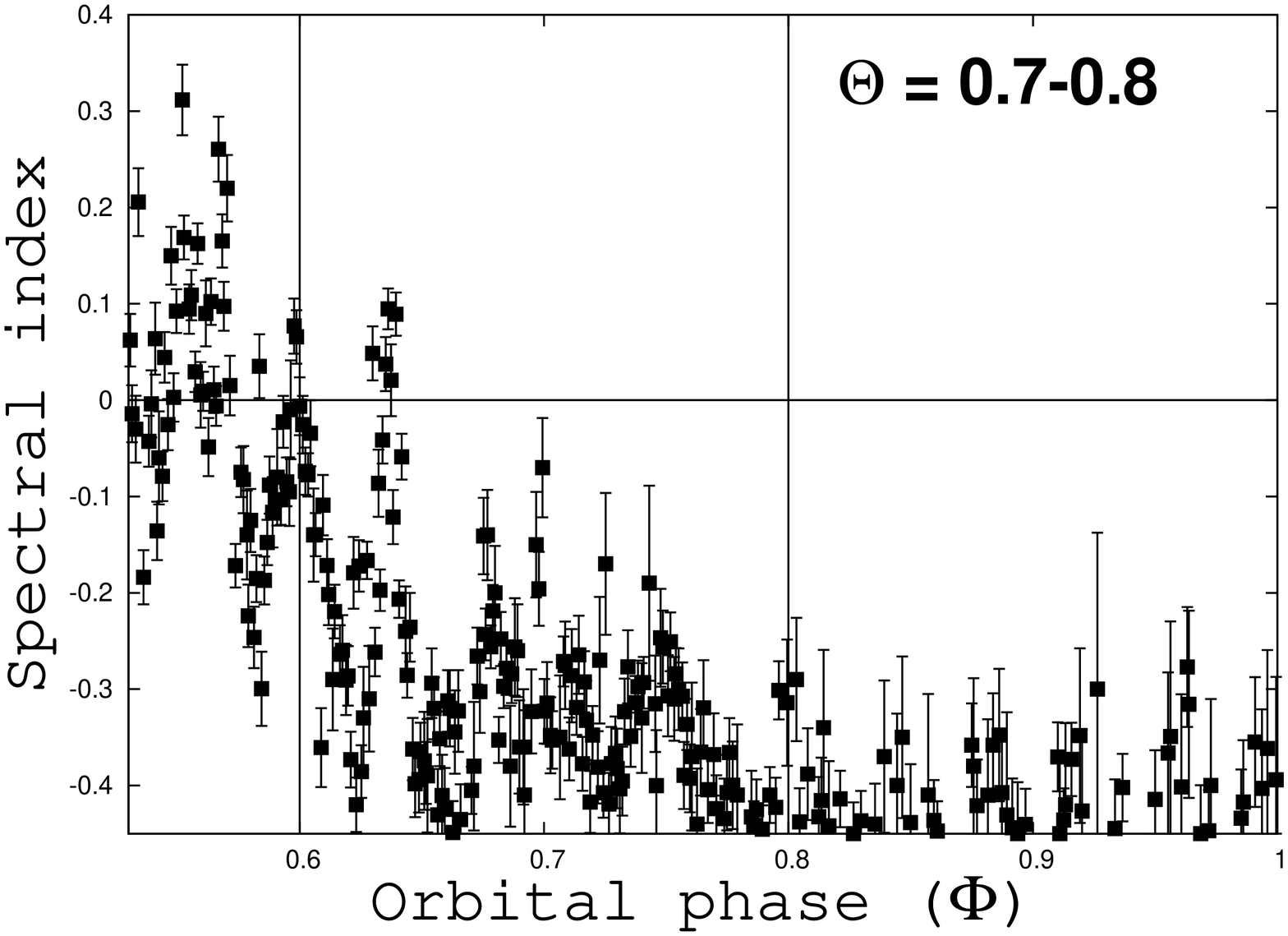}
      \caption{Radio spectral index for two different $\Theta$ intervals.
The vertical lines mark the $\Phi$ interval where \citet{Chernyakova06}
determined a photon index compatible with a steep power-law state (see Sect. 2).}
         \label{Fig3}
   \end{figure}

Radio spectral index curves showing $\Phi_{crit,2}$, the transition for the large radio outburst towards apastron, for $\Theta=$ 0.0-0.1 and 0.7-0.8 are shown in Fig. \ref{Fig3}. The orbital occurrence at phase $\Phi_{crit,2}$ of the transition from $\alpha\geq$ 0 to $\alpha<$ 0 lies for $\Theta=$ 0.0-0.1 around $\Phi_{crit}\approx$ 0.73-0.77, and for $\Theta=$ 0.7-0.8 it occurs around $\Phi_{crit}\approx$ 0.63-0.67. Owing to the connection between radio and X-ray states, it now becomes evident from Fig. \ref{Fig3} that, if the energy range of an instrument covers both the low hard and the steep power-law state, folding over the orbital period data of different $\Theta$s might result in the mixture of different states. The orbital phase of the transition from optically thick emission to an optically thin outburst does not remain the same over the 1667 d period and the unified X-ray state model with radio jets assumes a physical connection between the radio spectral transition and the X-ray states \citep{FenderBelloniGallo}. The folded data no longer provides any accurate information about the orbital occurrence of the spectral transition. As we show here, the information about one state could even be lost completely, if e.g. the sampling for one $\Theta$ interval is finer, which leads to the suppression of information from the other interval, and/or data from several different $\Theta$-intervals, and not only the two intervals shown in Fig. \ref{Fig3}, are mixed up. 

Folding all data or parts of I1, I2, and I3 together in terms of orbital phase increases the sampling and the significance of the signal. In doing so, several authors established that the emission in the interval 10-100 keV is clearly modulated by the orbital phase \citep{Chernyakova06,HermsenKuiper07,Zhang10}. Moreover, \citet{Chernyakova06}, who used data from almost all of I3 (covering $\Theta=0.57-0.99)$ and a few data points from the beginning of I1 (covering $\Theta=0.0-0.06$), have found that whereas along most of the orbit the photon index $\Gamma$ was $\approx$ 1.4-1.7 (i.e. low hard state), it changed  to $\Gamma=3.6^{+1.6}_{-1.1}$ (i.e. compatible with a steep power-law state) in the orbital phase interval $\Phi=$ 0.6-0.8. This result is consistent with the radio spectral index analysis of \citet{MassiKaufman}, giving for the same $\Theta$-intervals of I3 the transition to an optically thin outburst (a transient jet) after $\Phi=0.6 \pm$0.1. In Fig. \ref{Fig3}, one sees that for $\Theta=$ 0.7 - 0.8 (which is part of I3) the optically thin emission dominates in the interval $\Phi=$ 0.6-0.8. Following the radio spectral index though, folding over too much of a $\Theta$ cycle could imply that, although the resulting light curves can establish the overall periodicity of the source at these energies, it mixes different spectral states. The analysis of \citet{Zhang10}, who used I1, I2, and I3 together (following the same data reduction method as \citet{Chernyakova06}), thereby covering both the minimum and the maximum amplitude interval of $\Theta$, give a hard photon index of $\Gamma\approx$ 1.4-1.9 along the whole orbit (average: 1.7), no longer finding evidence of a very soft spectrum ($\Gamma \geq$ 2.4) in the interval $\Phi=$ 0.6-0.8. 

\begin{figure}
   \centering
\includegraphics[width=.32\textheight]{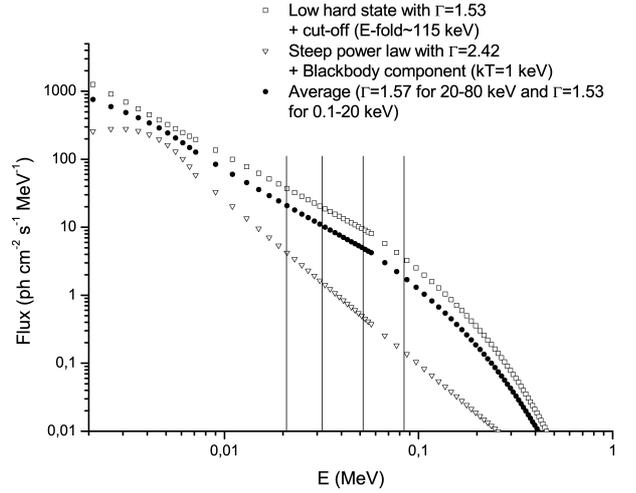}
      \caption{Model spectra for the low/hard state (power-law plus cut-off, open squares), the steep power-law state (open triangles) together with a black-body disk component with kT$\sim$1 keV and the average of the two states (filled circles). }
         \label{Fig6}
   \end{figure}

How can these results be interpreted in terms of the radio spectral index
analysis from above? By examining Fig. \ref{Fig3}, we can see, how by folding together data of two $\Theta$ intervals, optically thin and thick emission can be mixed, because the orbital occurrences of $\Phi_{crit,2}$, and therefore of the transition between optically thin/thick emission, differ for the $\Theta$ intervals. The optically thin emission then no longer dominates the orbital interval  $\Phi=$ 0.6-0.8. When averaging over the intervals I1, I2, and I3, optically thick and thin emission get mixed together in the $\Phi$ interval of interest. Therefore, the measured photon index must not be directly interpreted as a continuous low/hard state of the source, because it represents the
combination of different states. Moreover, we note that this mixing also occurs, when only one $\Theta$ interval is selected and the data is averaged over the whole orbit. As one sees in Fig. \ref{Fig3}, optically thick and thin emission occur alternately during the orbit. When averaging the spectrum across the whole orbit, \citet{Chernyakova06} find again an average photon index of $\Gamma\approx$ 1.6 but no cut-off at high energies. 
An example of the effect of this mixing of X-ray states is shown in Fig. \ref{Fig6}; here we show one spectrum for a low/hard state (open squares) and one spectrum for a steep power-law state (open triangles) and the spectrum for the low/hard state is
\begin{equation}
f(E)=\left\{ \begin{array}{rl}
E^{-\Gamma} & E<E_c\\
E^{-\Gamma} \exp{-(E-E_c)/E_{fold}} & E>E_c,
\end{array}\right.
\end{equation}
where $E_c$ is the cut-off energy and $E_{fold}$ the folding energy, which is in \citet{Grove} assumed to be approximately twice $E_c$. We use as an example the source GRS 1716-249 with $E_{fold}=115\pm 8$ keV, because this source, analysed by \citet{Grove} in both spectral states, results in a steep power-law state of significantly lower luminosity. This might also hold for \lsp, as suggested by the results of \citet{Chernyakova06} (i.e. dominance of the low hard state when mixing the data). In Fig. \ref{Fig6}, we superimpose an average of the two curves (filled circles). The four vertical lines indicate the four points of the averaged (over both $\Phi$ and $\Theta$) spectra derived by \citet{Chernyakova06}. The fit to these points resulted in a photon index $\Gamma$ consistent with a low hard state, but no cut-off at high energies was found. This example shows how an average could yield a curve, where the photon index of the power-law below the cut-off energy ($E_c\approx 115/2$ keV) is consistent with a low/hard state although the steep power-law is part of the data. In addition, we see in this example for GRS 1716-249 that only the last point, among the four ones by \citet{Chernyakova06}, samples the average curve after the cut-off and lies, with around 84 keV, still quite close to the cut-off energy ($E_c\approx 115/2$ keV). It would be difficult under this condition to detect the cut-off. For GRS 1716-249, Grove et al. indeed established that there is exponential decay with $E_{fold}=115\pm 8$keV for very well-sampled data up to 300 keV. However, if the cut-off in \lsi were to occur well before the assumed one, the contribution of the steep power-law would strongly bias the analysis of the averaged data.

\section{Conclusions and discussion}
We have investigated the implications of the radio spectral index transition,  $\Phi_{crit}=f(\Theta)$, in \lsi for the analysis of INTEGRAL data. In microquasars, this transition is directly linked to the transition between two spectrally distinct high energy states that cover the energy range from X-rays to gamma-rays \citep{Grove}, but which in this paper are simply referred to as X-ray states: the low hard X-ray state and the steep power-law X-ray state. For the energy range covered by INTEGRAL data, both states can be detected, when they occur. From our analysis, we conclude that:
\begin{itemize}
    \item Folding data over too large $\Theta$ intervals could mix up data from different states so that the resulting spectra can no longer be unambiguously interpreted and mimic a persistent low hard state. 
\item For data from the energy range accessible to INTEGRAL, even if the data were not folded on $\Theta$, averaging across $\Phi$ will destroy any information about the different states. 
\item We have pointed out that the lack of a cut-off in the power-law for data averaged all over the orbit, means that if \lsi has a cut-off well below $\sim$~115/2 keV, the contribution of the steep power-law will strongly bias the analysis of the spectrum of the averaged data. However, we have shown that a higher cut-off requires highly sensitive sampling, extending well beyond the used range of up to 100 keV.
\end{itemize}
The INTEGRAL results are then explainable with two X-ray states alternating along the orbit in parallel to, as expected in the microquasar scenario, the two radio states. What does this imply for the analysis of the HE/VHE data? For VHE, when \lsi is detected by MAGIC or VERITAS, its spectrum is always closely fitted with a photon index of $\gtrsim$2.4 \citep{Albert09,Acciari2011,Jogler2011}, which is the same as expected for the steep power-law. In this context, the observations of \citet{Anderhub09} are of special interest. They have observed simultaneously X-ray and VHE emission from \lsi and found a correlation. \citet{MassiZimmermann} noticed that for these observations the two fluxes result in $F_{\rm VHE}\propto F_{\rm X}^{\eta}$ with $\eta=0.99$, which is in agreement with the correlation observed in blazars \citep{Katarzynski2010}. Here the X-ray emission is due to synchrotron, and VHE is synchrotron self-Compton (SSC) \citep{Katarzynski2005}. For \lsp, several authors have explained the X-ray excess around apastron with synchrotron emission and the VHE with either external inverse Compton (EIC) or SSC emission \citep{Gupta2006,Zabalza2011}. \citet{Zabalza2011} demonstrated that the X-ray/VHE correlation observed by Anderhub et al. (2009) is compatible with a one-zone leptonic particle population producing the emission. Their models use the observed X-ray photon index (1.55-1.67) and can reproduce the observed lightcurves quite well. Interestingly, this X-ray photon index is comparable to the one expected for the low hard state and could be mistaken as such. The soft X-ray synchrotron emission during the steep power-law state can then explain two findings. First, the VHE emission by invoking EIC or SSC and, second, the absence of an accretion disk black-body component in the soft X-rays (0.1-10 keV) expected during the steep power-law state, if the synchrotron X-ray component outshines the disk. Owing to the lack of simultaneous hard X-ray/TeV observations of \lsp, it remains open, if the jet component extends into the hard X-ray regime. 

For the high energy emission in the GeV range, detected e.g. with \textit{Fermi}-LAT, the picture is more diverse. \citet{Zabalza2011} note that in their one-zone population model, a broken power-law electron energy distribution is needed to explain the GeV component as observed with \textit{Fermi}-LAT. The steep power-law might not then be the only emission process, also because GeV emission is detected all along the orbit. Electrons from the steady jet can always upscatter stellar UV photons to GeV energies (i.e. EIC see Bosch-Ramon et al. 2006). Nevertheless, more energetic particles from the transient jet could in addition produce as well GeV emission via EIC and SSC. Intriguingly, the spectrum measured by \textit{Fermi}-LAT shows, in addition to a power-law with a cut-off around 6 GeV, upper limits possibly compatible with the spectrum measured with MAGIC and VERITAS (see e.g. Fig. 2 in Hadasch 2011).

In any case, owing to the connection between the radio and the HE/VHE emission, the analysis performed here for INTEGRAL can and should in principle be extended to other high energy instruments, if the $\Theta$ coverage is good enough. In particular, the flux evolutions recently detected at HE/VHE imply that this data should be analysed on as short a $\Theta$ interval as possible. As discussed in \citet{MassiKaufman}, the two peak shape of the $\alpha$ versus $\Phi$ curve varies with $\Theta$. It is most pronounced in $\Theta$=0.0-0.1 and 0.9-1.0, appreciable at $\Theta$=0.2-0.3 and 0.7-0.9, and absent at $\Theta$=0.3-0.6 (see Fig. 5 in Massi \& Kaufman Bernad{\'o} 2009). In addition, the distance of the two peaks varies with $\Theta$. These variations in the accretion curve are predicted by the two-peak microquasar model of \citet{MartiParedes} by incorporating variable wind velocities of the Be star (see their Fig.6). A reflection of this behaviour at other wavelengths owing to the related emission processes is expected. As a matter of fact, there is an interesting increase in the overall flux level observed with \textit{Fermi}-LAT after March 2009 ($\Theta\approx$ 0.92) together with a broadening of the peak shape \citep{Hadasch2010}. The flux variations observed with \textit{Fermi}-LAT to date are therefore consistent with $\alpha$ varying with $\Theta$. Similarly, strong variations are observed at very high energies. The source went from being detected around apastron with VERITAS and MAGIC (VERITAS: $\Theta$=0.38, 0.4-0.47; MAGIC: $\Theta$=0.17-0.25, 0.37-0.44, 0.59-0.6) to becoming quiescent during 2008-2010. No detection was reported by VERITAS for $\Theta$=0.83-0.08, but weak detection was detected by MAGIC for $\Theta$=0.05-0.11 around apastron. The source was again detected by VERITAS in October 2010 corresponding to $\Theta$=0.26, but this time around periastron \citep{Albert09,Acciari2011,Jogler2011}. Nonetheless, at the moment the insufficient $\Theta$ coverage of these instruments does not allow a closer comparison. Nevertheless, the observations clearly indicate that for the analysis of these data, $\Theta$ should be considered to avoid any mixing of signal from the outbursts.

\acknowledgements
The work of L. Zimmermann is partly supported by the German Excellence
Initiative via the Bonn Cologne Graduate School. We thank Lars Fuhrmann, Victoria Grinberg and the anonymous referee for comments and fruitful discussions. The Green Bank Interferometer is a facility of the National Science Foundation operated by the NRAO in support of NASA High Energy Astrophysics programs.

\bibliographystyle{aa}

\begin{thebibliography}{}

\bibitem[Abdo et al.(2009)]{Abdo} Abdo, A.~A., et al.\ 2009, \apj, 701,
L123

\bibitem[Acciari et al.(2011)]{Acciari2011} Acciari, V.~A., et al.\ 
2011a, arXiv:1105.0449 

\bibitem[Albert et al.(2009)]{Albert09} Albert, J., et al.\ 2009, \apj,
693, 303


\bibitem[Anderhub \& al.(2009)]{Anderhub09} Anderhub, H., et al.\ 2009,
\apjl, 706, L27

\bibitem[Aragona et al.(2009)]{Aragona09} Aragona, C., McSwain, M.~V.,
Grundstrom, E.~D. et al. \ 2009, \apj, 698, 514


\bibitem[Bosch-Ramon et al.(2006)]{BoschRamon} Bosch-Ramon, V. et al. \ 2006, \aap, 459, L25

\bibitem[Caballero-Garc{\'{\i}}a et al.(2009)]{Caballero09}
Caballero-Garc{\'{\i}}a, M.~D., Miller, J.~M., Trigo et al. \ 2009,
\apj, 692, 1339

\bibitem[Casares et al.(2005)]{Casares} Casares, J. et al.\ 2005, \mnras, 360, 1105

\bibitem[Chernyakova et al.(2006)]{Chernyakova06} Chernyakova, M.,
Neronov, A., \& Walter, R.\ 2006, \mnras, 372, 1585

\bibitem[Connors et al.(2002)]{connors02} Connors, T.~W. et al.\ 2002, \mnras, 336, 1201 

\bibitem[Corbel et al.(2006)]{Corbel06} Corbel, S., Tomsick,
J.~A., \& Kaaret, P.\ 2006, \apj, 636, 971

\bibitem[Corbel et al.(2008)]{Corbel08} Corbel, S., Koerding,
E., \& Kaaret, P.\ 2008, \mnras, 389, 1697

\bibitem[Dhawan et al. (2006)]{Dhawan06} Dhawan, V.,  Mioduszewski, A., \&
Rupen, M. 2006, Proceedings of  the VI Microquasar Workshop, p. 52.1

\bibitem[Dubus(2006)]{Dubus} Dubus, G. \ 2006, \aap, 456, 801



\bibitem[Fender et al.(2004)]{FenderBelloniGallo} Fender, R.~P., Belloni,
T.~M., \& Gallo, E.\ 2004, \mnras, 355, 1105


\bibitem[Gregory et al.(1999)]{Gregory99} 
Gregory, P.~C., 
Peracaula, M., \& Taylor, A.~R.\ 1999, \apj, 520, 376 

\bibitem[Gregory(2002)]{Gregory02} Gregory, P.~C.\ 2002, \apj, 575, 427



\bibitem[Grove et al.(1998)]{Grove} Grove, J.~E., Johnson, W.~N., Kroeger,
R.~A. et al. \ 1998, \apj, 500, 899



\bibitem[Gupta \& Boettcher(2006)]{Gupta2006} Gupta, S. \& Boettcher, M.\ 2006, \apjl, 650, L123 

\bibitem[Hadasch(2011)]{Hadasch2010} Hadasch, D., for the Fermi-LAT collaboration 2011, [{\tt arXiv:1111.0350}]


\bibitem[Hermsen \& Kuiper(2007)]{HermsenKuiper07} Hermsen, W., \& Kuiper,
L. \ 2007, in First GLAST Symposium


\bibitem[Katarzy{\'n}ski et al.(2005)]{Katarzynski2005} Katarzy{\'n}ski, K., et al. 2005, \aap, 433, 479 

\bibitem[Katarzy\`nski \& Walczewska(2010)]{Katarzynski2010} Katarzy\`nski, K., Walczewska, K. \ 2010, \aap, 510, A63 

\bibitem[MAGIC Collaboration et al.(2011)]{Jogler2011} MAGIC 
Collaboration, Aleksi{\'c}, J., Alvarez, E.~A., et al.\ 2011, 
arXiv:1111.6572 

\bibitem[Maraschi \& Treves(1981)]{Maraschi81} Maraschi, L., Treves, A.\
1981, \mnras, 194, 1P



\bibitem[Marti \& Paredes(1995)]{MartiParedes} Marti, J., \& Paredes,
J.~M.\ 1995, \aap, 298, 151

\bibitem[Massi et al.(2004)]{massi04} Massi, M. et al. \ 2004, \aap, 414, L1-L4


\bibitem[Massi \& Kaufman Bernad{\'o}(2009)]{MassiKaufman} Massi, M., \&
Kaufman Bernad{\'o}, M.\ 2009, \apj, 702, 1179

\bibitem[Massi \& Zimmermann(2010)]{MassiZimmermann} Massi, M., \&
Zimmermann, L. \ 2010, \aap, 515, A82


\bibitem[Massi(2011a)]{Massi10a} Massi, M.\ 2011a,  Mem. Soc. Astron. It., 82, 24 


\bibitem[Massi(2011b)]{Massi10b} Massi, M.\ 2011b, Mem. Soc. Astron. It., 82, 77


\bibitem[McClintock \& Remillard(2006)]{McClintockRemillard06} McClintock,
J.~E., \& Remillard, R.~A.\ 2006, Compact Stellar X-ray Sources, Cambridge University Press, p. 157


\bibitem[Romero et al.(2007)]{Romero2007} Romero, G.~E. et al. \ 2007, \aap, 474, 15

\bibitem[Sidoli et al.(2006)]{Sidoli06} Sidoli, L., Pellizzoni, A.,
Vercellone, S. et al. \ 2006, \aap, 459, 901


\bibitem[Zabalza et 
al.(2011)]{Zabalza2011} Zabalza, V., Paredes, J.~M., \& Bosch-Ramon, V.\ 2011, \aap, 527, A9 


\bibitem[Zhang et al.(2010)]{Zhang10} Zhang, S., Torres, D.~F., Li, J. et al.
\ 2010, \mnras, 408, 642

\bibitem[Zimmermann et al.(2011)]{Zimmermann} Zimmermann, L., 
Grinberg, V., Massi, M., \& Wilms, J.\ 2011, The X-ray Universe 2011, 308 

\end{thebibliography}

\end{document}